\documentclass[12pt]{article}
\usepackage[margin=1.9cm]{geometry}
\usepackage{graphicx}
\usepackage{cite}
\usepackage{xcolor}
\usepackage{amsmath,amssymb}

\newcommand{\mysection}{\setcounter{equation}{0}\section}

\def\beq{\begin{equation}}
\def\eeq{\end{equation}}
\def\beqa{\begin{eqnarray}}
\def\eeqa{\end{eqnarray}}
 
\begin{document}
\begin{center}
{\Large \bf Higher-order soft and virtual corrections in $p p \to \gamma W$ production at the LHC}
\end{center}

\vspace{2mm}

\begin{center}
{\large Nikolaos Kidonakis$^a$ and Alberto Tonero$^{a,b}$} \\

\vspace{2mm}

${}^a${\it Department of Physics, Kennesaw State University, \\
Kennesaw, GA 30144, USA}

\vspace{1mm}

${}^b${\it Department of Chemistry and Physics, Florida Gulf Coast University, \\
Fort Myers, FL 33965, USA}

\end{center}

\begin{abstract}
We study higher-order QCD corrections beyond NLO for the associated production of a photon with a $W$ boson ($\gamma W^+$ and $\gamma W^-$ production) at the Large Hadron Collider. We calculate the NNLO soft-plus-virtual QCD corrections as well as the N$^3$LO soft-gluon corrections to the total production cross section and the photon transverse-momentum distribution in single-particle-inclusive kinematics. The higher-order corrections provide a significant enhancement to the cross section. This is the first calculation of the complete soft-gluon corrections at N$^3$LO in single-particle-inclusive kinematics for a Standard Model process.
\end{abstract}

\mysection{Introduction}

The associated production of a photon with a $W$ boson ($\gamma W^+$ and $\gamma W^-$) has been studied experimentally and theoretically for a long time at relevant collider energies. The interest in these processes is in part due to the possibility of probing the triple gauge coupling of a photon with two $W$ bosons since such diagrams already contribute at lowest order in addition to diagrams with photon and $W$ couplings to quark lines. Thus, measurements of the $\gamma W$ cross section can be used to look for effects of new physics and set limits on anomalous gauge couplings.

The $\gamma W$ production cross section was first measured in 1.8 TeV \cite{CDF1.8,D01.8a,D01.8b} and 1.96 TeV \cite{CDF1.96,D01.96a,D01.96b,D01.96c} proton-antiproton collisions at the Tevatron. It was later studied in proton-proton collisions at the LHC at 7 TeV \cite{CMS7a,ATLAS7a,ATLAS7b,ATLAS7c,CMS7b}, 8 TeV \cite{ATLAS8}, and 13 TeV \cite{CMS13a,CMS13b,ATLAS13} energies.

Theoretical studies for $\gamma W$ production have appeared at various orders in Refs. \cite{LO,NLO1,NNLOsv,NLOhel,EW1,NLOg,EW2,NLOdi,NLOew,NNLO}. The leading-order (LO) calculation was done in \cite{LO}, the next-to-leading-order (NLO) QCD corrections were calculated in \cite{NLO1}, the next-to-next-to-leading-order (NNLO) soft-plus-virtual QCD corrections were calculated in \cite{NNLOsv}, and the NNLO QCD corrections were presented in \cite{NNLO}.

Soft-gluon corrections are important for a wide variety of processes in the Standard Model and beyond because the production cross sections receive large contributions from soft-gluon emission near partonic threshold. In this work, we calculate the soft-gluon corrections through next-to-next-to-next-to-leading-order (N$^3$LO) based on the general resummation formalism in Refs. \cite{GS87,CT89,NKGS1,NKGS2,LOS,NKVD,NKASV,GKSV,NKhiggs,NKtW,NKRJG,NKWH,NKAT} and using single-particle-inclusive kinematics with the photon as the identified particle. We present approximate NNLO (aNNLO) and approximate N$^3$LO (aN$^3$LO) total and differential cross sections to $\gamma W$ production. The aNNLO theoretical predictions are derived by adding the complete NNLO soft-plus-virtual corrections to the exact NLO QCD result. The aN$^3$LO predictions are derived by further adding the complete N$^3$LO soft-gluon corrections to the aNNLO results. 

The paper is organized as follows. In Section 2, we describe the soft-gluon resummation formalism for $\gamma W$ partonic processes and present detailed analytical results for the soft-gluon corrections through N$^3$LO. In Section 3, we present numerical results for the total cross section through aN$^3$LO at 13 TeV and 13.6 TeV collision energies at the LHC. We provide separate reults for $\gamma W^+$ and $\gamma W^-$ production and give uncertainties from scale variation and from parton distributions. In Section 4, we present results for the photon differential distributions in transverse momentum at 13 and 13.6 TeV LHC energies. We conclude in Section 5.

\mysection{Soft-gluon corrections for $\gamma W$ production}

In this section, we discuss the resummation formalism that we use for the calculation of soft-gluon corrections in $\gamma W$ production. The origin of the soft-gluon corrections is from the emission of soft (low-energy) gluons, which result in partial cancellations of infrared divergences between real-emission and virtual diagrams close to partonic threshold for the production of a $\gamma W$ pair.

In this work we consider $\gamma W$ production in proton-proton collisions, which occurs at leading-order through the following partonic processes  
\beq
q (p_a)\, + {\bar q'} (p_b) \to \gamma (p_1) \, + \, W (p_2)  \, ,
\eeq
where $q$ and ${\bar q'}$ are quarks and antiquarks in the protons, and $W$ denotes either a $W^+$ or a $W^-$ boson. We work in single-particle-inclusive kinematics where the photon is the observed particle. We define $s=(p_a+p_b)^2$, $t=(p_a-p_1)^2$, $t_1=t-m_W^2$, $u=(p_b-p_1)^2$, $u_1=u-m_W^2$, as well as a partonic threshold variable $s_4=(p_2+p_g)^2-m_W^2=s+t+u-m_W^2$, with $m_W$ the $W$-boson mass and $p_g$ the momentum of an additional gluon in the final state. Near partonic threshold $p_g \to 0$ and, thus, $s_4 \to 0$. 
The soft-gluon corrections appear in the perturbative series in the strong coupling, $\alpha_s$, as logarithms of $s_4$, i.e. $[(\ln^k(s_4/m_W^2))/s_4]_+$, with $0 \le k \le 2n-1$ at $n$th order in $\alpha_s$.
 
The resummation of soft-gluon corrections follows as a consequence of the factorization properties of the (differential) cross section under Laplace transforms. We express the differential hadronic cross section, $d\sigma_{pp \to \gamma W}$, as a convolution in the form 
\beq
d\sigma_{pp \to \gamma W }=\sum_{q,{\bar q'}} \; 
\int dx_a \, dx_b \,  \phi_{q/p}(x_a, \mu_F) \, \phi_{{\bar q'}/p}(x_b, \mu_F)  \, 
d{\hat \sigma}_{q{\bar q'} \to \gamma W}(s_4, \mu_F)  \, ,
\label{factorized}
\eeq
where $d{\hat \sigma}_{q{\bar q'} \to \gamma W}$ is the differential partonic cross section at factorization scale $\mu_F$, while $\phi_{q/p}$ and $\phi_{{\bar q'}/p}$ are parton distribution functions (pdf) with $x_a$, $x_b$ the momentum fractions of partons $q$, ${\bar q'}$ in the two protons. 

The LO differential partonic cross section is given by
\beq
\frac{d{\hat{\sigma}}^{(0)}_{q{\bar q'} \to \gamma W}}{dt_1 \, du_1}  = F^{(0)}_{q{\bar q'} \to \gamma W} \, \delta(s_4)
\label{LO}
\eeq
where
\beqa
F^{(0)}_{q{\bar q'} \to \gamma W}&=&\frac{\alpha^2 \, \pi \, m_Z^2}{54 s^2 (m_Z^2-m_W^2)}
\frac{(t-2u)^2(m_W^4+s^2-2\,t\,u)}{(s-m_W^2)^2\,t\,u}
\eeqa 
with $\alpha$ the electromagnetic coupling and $m_Z$ the $Z$-boson mass. 

The resummed cross section is given under Laplace transforms, with $N$ the transform variable, by 
\beqa
d\hat{\sigma}^{res}_{q{\bar q'} \to \gamma W}(N)&=&   
\exp\left[E_q (N_a)+ E_{\bar q'} (N_b)\right] \; 
\exp \left[2\int_{\mu_F}^{\sqrt{s}} \frac{d\mu}{\mu}\; 
\left(\gamma_{q/q}\left({\tilde N}_a,\alpha_s(\mu)\right)
+\gamma_{{\bar q'}/{\bar q'}}\left({\tilde N}_b,\alpha_s(\mu)\right)\right)
\right]
\nonumber \\ && \quad
\times \,\, H_{q{\bar q'} \to \gamma W}\left(\alpha_s({\sqrt s})\right) \;
{\tilde S}_{q{\bar q'} \to \gamma W}\left(\alpha_s({\sqrt s}/N)\right) \, .
\label{resdxs}
\eeqa

In the first exponent of Eq.~\eqref{resdxs},
\beq
E_q(N_a)=
\int^1_0 dz \frac{z^{N_a-1}-1}{1-z}\;
\left \{\int_1^{(1-z)^2} \frac{d\lambda}{\lambda}
A_q \left(\alpha_s(\lambda s)\right)
+D_q \left[\alpha_s((1-z)^2 s)\right]\right\} 
\label{Eexp}
\eeq
with $N_a=N(-u_1/s)$, and a similar expression holds for $E_{\bar q'}(N_b)$ with $N_b=N(-t_1/s)$. The first integrand in Eq.~\eqref{Eexp} admits the perturbative expansion 
$A_q = \sum_{n=1}^{\infty} (\alpha_s/\pi)^n A_q^{(n)}$.
The first term in this expansion is $A_q^{(1)}=C_F=(N_c^2-1)/(2N_c)$ \cite{GS87}, where $N_c=3$ is the number of colors. The second term, $A_q^{(2)}$, is given by \cite{CT89}
\beq
A_q^{(2)} =C_F C_A \left(\frac{67}{36}-\frac{\zeta_2}{2}\right)-\frac{5}{18}C_F n_f\,,
\eeq         
where $\zeta_2=\pi^2/6$, $C_A=N_c$, and $n_f=5$ is the number of light-quark flavors. Finally, the third term $A_q^{(3)}$ is given by \cite{MVV04}
\beqa
A_q^{(3)}&=&C_F C_A^2\left(\frac{245}{96}-\frac{67}{36}\zeta_2
+\frac{11}{24}\zeta_3+\frac{11}{8}\zeta_4\right)
+C_F^2 n_f\left(-\frac{55}{96}+\frac{\zeta_3}{2}\right)
\nonumber \\ &&
{}+C_F C_A n_f \left(-\frac{209}{432}+\frac{5}{18}\zeta_2
-\frac{7}{12} \zeta_3\right)-C_F\frac{n_f^2}{108} 
\eeqa
where $\zeta_3=1.202056903\cdots$ and $\zeta_4=\pi^4/90$.  
The second integrand in Eq.~\eqref{Eexp} has the perturbative expansion $D_q=\sum_{n=1}^{\infty} (\alpha_s/\pi)^n D_q^{(n)}$. 
In the Feynman gauge, the one-loop term $D_q^{(1)}=0$, while
at two loops \cite{CLS97}
\beq
D_q^{(2)}=C_F C_A \left(-\frac{101}{54}+\frac{11}{6} \zeta_2
+\frac{7}{4}\zeta_3\right)
+C_F n_f \left(\frac{7}{27}-\frac{\zeta_2}{3}\right) \, ,
\eeq
and at three loops \cite{MA05}
\beqa
D_q^{(3)}&=&C_F C_A^2 \left(-\frac{297029}{46656}
+\frac{6139}{648} \zeta_2+\frac{2509}{216} \zeta_3
-\frac{187}{48} \zeta_4 -\frac{11}{12} \zeta_2 \zeta_3-3 \zeta_5\right)
\nonumber \\ && 
{}+C_F C_A n_f \left(\frac{31313}{23328}-\frac{1837}{648}\zeta_2
-\frac{155}{72}\zeta_3+\frac{23}{24}\zeta_4\right)
\nonumber \\ && 
{}+C_F^2 n_f \left(\frac{1711}{1728}-\frac{\zeta_2}{4}-\frac{19}{36}\zeta_3
-\frac{\zeta_4}{4}\right)
+C_F n_f^2 \left(-\frac{29}{729}+\frac{5}{27}\zeta_2
+\frac{5}{54}\zeta_3 \right)
\eeqa
where $\zeta_5=1.036927755\cdots$.

In the second exponent of Eq.~\eqref{resdxs}, the quantity $\gamma_{q/q}$ is the moment-space anomalous dimension of the ${\overline {\rm MS}}$ density $\phi_{q/q}$ \cite{FRS1,FRS2,GALY79,GFP80,FP80} (and similarly for  $\gamma_{{\bar q'}/{\bar q'}}$), and it can be expressed as 
$\gamma_{q/q}({\tilde N})=-A_q \ln {\tilde N} +\gamma_q$ where ${\tilde N}=N e^{\gamma_E}$ with $\gamma_E$ the Euler constant. The perturbative expansion of the parton anomalous dimension is
$\gamma_q=\sum_{n=1}^{\infty} (\alpha_s/\pi)^n \gamma_q^{(n)}$
with $\gamma_q^{(1)}=3C_F/4$ and
\beq
\gamma_q^{(2)}=C_F^2\left(\frac{3}{32}-\frac{3}{4}\zeta_2
+\frac{3}{2}\zeta_3\right)
+C_F C_A\left(\frac{17}{96}+\frac{11}{12}\zeta_2-\frac{3}{4}\zeta_3\right)
-C_F n_f \left(\frac{1}{48}+\frac{\zeta_2}{6}\right)\, .
\eeq

We expand Eq.~\eqref{resdxs} to fixed order and then invert back to momentum space to derive results for the physical cross section.
We further define
\beq
{\cal D}_k(s_4)=\left[\frac{\ln^k(s_4/m_W^2)}{s_4}\right]_+ \,.
\eeq

The NLO soft-plus-virtual $(S+V)$ corrections to the partonic cross section are given by
\beq
\frac{d{\hat{\sigma}}^{(1) \, S+V}_{q{\bar q'} \to \gamma W}}{dt_1 \, du_1} = F^{(0)}_{q{\bar q'} \to \gamma W}
\frac{\alpha_s(\mu_R)}{\pi}
\left\{c_3\, {\cal D}_1(s_4) + c_2\,  {\cal D}_0(s_4) 
+c_1\,  \delta(s_4)\right\}
\label{NLO}
\eeq
where $\mu_R$ is the renormalization scale, $c_3=4 C_F$,
\beq
c_2= -2 \, C_F \, \ln\left(\frac{t_1 u_1}{m_W^4}\right)
-2 \, C_F \, \ln\left(\frac{\mu_F^2}{s}\right) \, ,
\label{c3c2}
\eeq
and
\beq
c_1= C_F\, \ln^2\left(\frac{-t_1}{m_W^2}\right)
+ C_F\, \ln^2\left(\frac{-u_1}{m_W^2}\right)
+\left[C_F \, \ln\left(\frac{t_1 u_1}{m_W^4}\right)-2\gamma_q^{(1)}\right]
\ln\left(\frac{\mu_F^2}{s}\right)+V_1\,,
\eeq
where $V_1=2C_F(-2+\zeta_2)$ denotes the contribution of the one-loop virtual corrections \cite{KAP,AEM,HvN}. 

The NNLO soft-plus-virtual corrections are given by
\beq
\frac{d{\hat{\sigma}}^{(2) \, S+V}_{q{\bar q'} \to \gamma W}}{dt_1 \, du_1} = F^{(0)}_{q{\bar q'} \to \gamma W} \frac{\alpha_s^2(\mu_R)}{\pi^2} \left\{\sum_{k=0}^3 C^{(2)}_k \, D_k(s_4) + C^{(2)}_{\delta} \delta(s_4) \right\}
\label{NNLO}
\eeq
where
\beqa
C^{(2)}_3 &=& \frac{1}{2} c_3^2 \, ,
\nonumber \\
C^{(2)}_2 &=& \frac{3}{2}c_3 c_2-\frac{\beta_0}{4} c_3 \, ,
\nonumber \\ 
C^{(2)}_1 &=& c_3 c_1+c_2^2-\zeta_2 c_3^2
-\frac{\beta_0}{2} c_2+4 A_q^{(2)}
-\beta_0 C_F\ln\left(\frac{\mu_F^2}{\mu_R^2}\right) \, ,
\nonumber \\ 
C^{(2)}_0 &=& c_2 c_1-\zeta_2 c_3 c_2+\zeta_3 c_3^2+2 D_q^{(2)}
-\frac{\beta_0}{2} C_F \left[\ln^2\left(\frac{-t_1}{m_W^2}\right)+\ln^2\left(\frac{-u_1}{m_W^2}\right)\right]        
\nonumber \\ &&
{} -2 A_q^{(2)} \ln\left(\frac{t_1 u_1}{m_W^4}\right)
+\frac{\beta_0}{4} c_2 \ln\left(\frac{\mu_R^2}{s}\right)
-2 A_q^{(2)} \ln\left(\frac{\mu_F^2}{s}\right)
+\frac{\beta_0}{4} C_F \ln^2\left(\frac{\mu_F^2}{s}\right) \, ,
\nonumber \\ 
C^{(2)}_{\delta} &=& V_2+\frac{1}{2} \left(c_1^2-V_1^2\right)-\frac{\zeta_2}{2}c_2^2
+\zeta_3 c_3 c_2 +\frac{\beta_0}{6} C_F \left[\ln^3\left(\frac{-t_1}{m_W^2}\right)+\ln^3\left(\frac{-u_1}{m_W^2}\right)\right]
\nonumber \\ && 
{}+\left(\frac{\beta_0}{4} C_F+A_q^{(2)}\right) \left[\ln^2\left(\frac{-t_1}{m_W^2}\right)+\ln^2\left(\frac{-u_1}{m_W^2}\right)\right]
+\frac{\beta_0}{4} c_1 \ln\left(\frac{\mu_R^2}{s}\right)
-2 \gamma_q^{(2)} \ln\left(\frac{\mu_F^2}{s}\right)
\nonumber \\ && 
{}+A_q^{(2)} \ln\left(\frac{t_1 u_1}{m_W^4}\right) \ln\left(\frac{\mu_F^2}{s}\right)
+\frac{\beta_0}{8} \left[2 \gamma_q^{(1)}-C_F \ln\left(\frac{t_1 u_1}{m_W^4}\right)\right]
\ln^2\left(\frac{\mu_F^2}{s}\right) \, , 
\label{NNLOcoeff}
\eeqa
with $\beta_0=11 C_A/3-2n_f/3$ denoting the lowest-order $\beta$ function \cite{GW,HDP,tHooft}
and
\beqa
V_2&=&C_F^2\left(\frac{511}{64}-\frac{35}{8}\zeta_2-\frac{15}{4}\zeta_3
+\frac{\zeta_2^2}{10}\right)+C_F C_A\left(-\frac{1535}{192}
+\frac{37}{9}\zeta_2+\frac{7}{4}\zeta_3-\frac{3}{20}\zeta_2^2\right)
\nonumber \\ &&
{}+C_F n_f \left(\frac{127}{96}-\frac{7}{9}\zeta_2+\frac{\zeta_3}{2}\right)
\eeqa
denoting the contribution of the virtual two-loop corrections \cite{V2qq}.
It is important to note that our above expressions for the soft-plus-virtual corrections at NLO and NNLO are in complete agreement with the analytical results presented in Ref. \cite{NNLOsv} though they are presented in a somewhat different format in \cite{NNLOsv} (see also the discussion in Section 3.4 of Ref. \cite{NKrev}). 

The N$^3$LO soft-gluon $(S)$ corrections are given by
\beq
\frac{d^2{\hat{\sigma}}^{(3) \, S}_{q{\bar q'} \to \gamma W}}{dt_1 \, du_1} = 
F^{(0)}_{q{\bar q'} \to \gamma W} \frac{\alpha_s^3(\mu_R)}{\pi^3} \, \sum_{k=0}^5 C^{(3)}_k \, D_k(s_4) 
\label{N3LO}
\eeq
where
\beqa
C^{(3)}_5 &=& \frac{1}{8} c_3^3 \, ,
\nonumber \\
C^{(3)}_4 &=& \frac{5}{8} c_3^2 c_2 -\frac{5}{24} c_3^2 \beta_0 \, ,
\nonumber \\
C^{(3)}_3 &=& c_3 c_2^2 +\frac{1}{2}c_3^2 c_1-\zeta_2 c_3^3 
+\frac{\beta_0^2}{12}c_3-\frac{5}{6}\beta_0 c_3 c_2+4 c_3 A_q^{(2)} 
-\beta_0 C_F c_3 \ln\left(\frac{\mu_F^2}{\mu_R^2}\right) \, ,
\nonumber \\
C^{(3)}_2 &=& \frac{3}{2} c_3 c_2 c_1 +\frac{1}{2} c_2^3
-3 \zeta_2 c_3^2 c_2 +\frac{5}{2} \zeta_3 c_3^3
-\frac{\beta_0}{4} c_3 c_1 +\frac{9}{8} \beta_0 \zeta_2 c_3^2 -C_F \frac{\beta_1}{4}
\nonumber \\ &&
{}+\left(3 c_2-\beta_0 \right) \left[-\frac{\beta_0}{4} c_2 + 2 A_q^{(2)}
-\frac{\beta_0}{2} C_F \ln\left(\frac{\mu_F^2}{\mu_R^2}\right) \right]
-\frac{3}{2} c_3 X_1 \, ,
\nonumber \\
C^{(3)}_1 &=& \frac{1}{2} c_3 c_1^2+c_2^2 c_1 -\zeta_2 c_3^2 c_1
-\frac{5}{2} \zeta_2 c_3 c_2^2+5 \zeta_3 c_3^2 c_2+\frac{5}{4} \zeta_2^2 c_3^3
-\frac{15}{4} \zeta_4 c_3^3 
\nonumber \\ &&         
{}-\frac{\beta_0^2}{4} \zeta_2 c_3
-\frac{5}{3} \beta_0 \zeta_3 c_3^2+\beta_0 \zeta_2 c_3 c_2+4 A_q^{(3)}
+C_F \frac{\beta_1}{4}\ln\left(\frac{t_1 u_1}{m_W^4}\right)
\nonumber \\ &&         
{}+\left(2 c_1-5 \zeta_2 c_3\right) \left[-\frac{\beta_0}{4}c_2 + 2 A_q^{(2)}
-\frac{\beta_0}{2} C_F \ln\left(\frac{\mu_F^2}{\mu_R^2}\right) \right]
+C_F \frac{\beta_1}{4} \ln\left(\frac{\mu_R^2}{s}\right)
\nonumber \\ && 
{}-2\beta_0 A_q^{(2)} \ln\left(\frac{\mu_F^2}{\mu_R^2}\right)
+C_F \frac{\beta_0^2}{4} \ln^2\left(\frac{\mu_F^2}{\mu_R^2}\right)
+(\beta_0 -2 c_2) X_1+c_3 X_0 \, ,
\nonumber \\
C^{(3)}_0 &=& \frac{1}{2} c_2 c_1^2+3 \zeta_5 c_3^3-\frac{15}{4} \zeta_4 c_3^2 c_2
-2 \zeta_2 \zeta_3 c_3^3+ \zeta_3 c_3^2 c_1+2 \zeta_3 c_3 c_2^2
+\frac{5}{4} \zeta_2^2 c_3^2 c_2 -\zeta_2 c_3 c_2 c_1-\frac{\zeta_2}{2} c_2^3 
\nonumber \\ && 
{}+\frac{\beta_0}{12} c_3 \left(15 \zeta_4 c_3-8 \zeta_3 c_2-6 \zeta_2^2 c_3+3 \zeta_2 c_1\right)+2 D_q^{(3)}-\frac{C_F}{6} \beta_0^2 \left[\ln^3\left(\frac{-t_1}{m_W^2}\right)+\ln^3\left(\frac{-u_1}{m_W^2}\right)\right]
\nonumber \\ && 
{}-\left[\frac{\beta_0}{4}\left(C_F \beta_0+4 A_q^{(2)}\right)+\frac{\beta_1}{8}C_F\right] \left[\ln^2\left(\frac{-t_1}{m_W^2}\right)+\ln^2\left(\frac{-u_1}{m_W^2}\right)\right] 
\nonumber \\ &&  
{}+\left(\beta_0 D_q^{(2)}-2A_q^{(3)}\right) \ln\left(\frac{t_1 u_1}{m_W^4}\right)
+\left(4 \zeta_3 c_3-3 \zeta_2 c_2\right) \left[-\frac{\beta_0}{4} c_2+2 A_q^{(2)}
-\frac{\beta_0}{2} C_F \ln\left(\frac{\mu_F^2}{\mu_R^2}\right)\right]
\nonumber \\ &&  
{}+D_q^{(2)} \beta_0 \ln\left(\frac{\mu_R^2}{s}\right)-2 A_q^{(3)} \ln\left(\frac{\mu_F^2}{s}\right)-\frac{\beta_0^2}{4} C_F \left[\ln^2\left(\frac{-t_1}{m_W^2}\right)
+\ln^2\left(\frac{-u_1}{m_W^2}\right)\right] \ln\left(\frac{\mu_R^2}{s}\right)
\nonumber \\ &&  
{}-\left(A_q^{(2)} \beta_0+C_F \frac{\beta_1}{8}\right) \ln\left(\frac{t_1 u_1}{m_W^4}\right) \ln\left(\frac{\mu_R^2}{s}\right)
+\frac{\beta_0^2}{16} c_2 \ln^2\left(\frac{\mu_R^2}{s}\right)
+\frac{\beta_0^2}{8} C_F \ln^2\left(\frac{\mu_F^2}{s}\right) \ln\left(\frac{\mu_R^2}{s}\right)
\nonumber \\ &&  
{}+\frac{1}{16}\left(C_F \beta_1+8\beta_0 A_q^{(2)}\right) \left[\ln^2\left(\frac{\mu_F^2}{\mu_R^2}\right)-\ln^2\left(\frac{\mu_R^2}{s}\right)\right]-C_F \frac{\beta_0^2}{24} \ln^3\left(\frac{\mu_F^2}{s}\right)
\nonumber \\ &&  
{}+\left(\zeta_2 c_3 -c_1 \right) X_1 + c_2 X_0 \, ,
\label{N3LOcoeff}
\eeqa
with
\beqa
X_1&=&\frac{\beta_0}{4} \zeta_2 c_3 -2 D_q^{(2)}  
+\frac{\beta_0}{2} C_F \left[\ln^2\left(\frac{-t_1}{m_W^2}\right)
+\ln^2\left(\frac{-u_1}{m_W^2}\right)\right]
+2 A_q^{(2)} \ln\left(\frac{t_1 u_1}{m_W^4}\right) 
\nonumber \\ &&
{}-\frac{\beta_0}{4} c_2 \ln\left(\frac{\mu_R^2}{s}\right)
+2 A_q^{(2)} \ln\left(\frac{\mu_F^2}{s}\right)
-\frac{\beta_0}{4} C_F \ln^2\left(\frac{\mu_F^2}{s}\right) \, , 
\label{X1}
\eeqa
\beqa
X_0&=& V_2-\frac{1}{2}V_1^2 +\frac{1}{20}\zeta_2^2 c_3^2 
+\frac{\beta_0}{6} \zeta_3 c_3-\frac{\beta_0}{4} \zeta_2 c_2 + 2 A_q^{(2)} \zeta_2
+\frac{\beta_0}{6} C_F \left[\ln^3\left(\frac{-t_1}{m_W^2}\right)
+\ln^3\left(\frac{-u_1}{m_W^2}\right)\right]
\nonumber \\ &&
{}+\left(C_F \frac{\beta_0}{4}+A_q^{(2)}\right)
\left[\ln^2\left(\frac{-t_1}{m_W^2}\right)
+\ln^2\left(\frac{-u_1}{m_W^2}\right)\right]
+\frac{\beta_0}{4} c_1 \ln\left(\frac{\mu_R^2}{s}\right)
-\frac{\beta_0}{2} \zeta_2 C_F \ln\left(\frac{\mu_F^2}{\mu_R^2}\right)
\nonumber \\ &&
{}+\left[-2 \gamma_q^{(2)} +A_q^{(2)} \ln\left(\frac{t_1 u_1}{m_W^4}\right)\right] 
\ln\left(\frac{\mu_F^2}{s}\right)
+\frac{\beta_0}{8} \left[ 2 \gamma_q^{(1)}-C_F \ln\left(\frac{t_1 u_1}{m_W^4}\right)\right] \ln^2\left(\frac{\mu_F^2}{s}\right) \, ,
\label{X0}
\eeqa
and $\beta_1=34 C_A^2/3-2 C_F n_f-10 C_A n_f/3$ \cite{WEC,DRTJ,ET79} denoting the coefficient of the second term in the perturbative expansion of the $\beta$ function.

\mysection{Total cross sections for $\gamma W^+$ and $\gamma W^-$ at 13 and 13.6 TeV}

In this section we present results for $\gamma W$ production at the LHC, for two center-of-mass energies, namely 13 TeV and 13.6 TeV. We present separate results for $\gamma W^+$ and $\gamma W^-$ total cross sections. The complete NLO results (including QCD and electroweak corrections) are calculated using {\small \sc MadGraph5\_aMC@NLO} \cite{MG5,MGew}. We use MSHT20 nnlo \cite{MSHT20nnlo} and an$^3$lo pdf \cite{MSHT20an3lo} in our calculations (we use lower-case letters for the order of the pdf, and capital letters for the order of the calculation of the perturbative cross section, in order to avoid confusion). We take $m_W=80.3692$ GeV, $m_Z=91.188$ GeV \cite{RPP24}, and $\alpha^{-1}=127.9$.  The central results are obtained by setting $\mu_F=\mu_R=m_W$. To obtain scale uncertainties, we vary the factorization and renormalization scales independently from each other over the range $m_W/2$ to $2m_W$, and we also compute pdf uncertainties. The aNNLO cross sections are calculated by adding second-order soft-plus-virtual QCD corrections to the complete NLO result. Third-order soft-gluon corrections are further added in order to derive a result at aN$^3$LO.

\begin{table}[htbp]
\begin{center}
\begin{tabular}{|c|c|c|c|c|c|} \hline
\multicolumn{6}{|c|}{$p p \to \gamma W^+$ and $p p \to \gamma W^-$ cross sections at 13 TeV \hspace{8mm} $p_{\gamma T}^{\rm cut}=25$ GeV} \\ \hline
process & pdf order &  $\sigma$ LO (pb) & $\sigma$ NLO (pb) & $\sigma$ aNNLO (pb) & $\sigma$ aN$^3$LO (pb)  \\ \hline
$\gamma W^+$ & nnlo & $23.6^{+1.7}_{-1.9}{}^{+0.4}_{-0.3}$ & $55.3^{+5.2}_{-5.1}{}^{+0.5}_{-0.6}$ & $60.4^{+4.6}_{-5.6}{}^{+0.5}_{-0.7}$ & $62.0^{+5.1}_{-5.9}{}^{+0.5}_{-0.7}$ \\ \hline
$\gamma W^+$ & an$^3$lo & $23.8^{+1.6}_{-1.8}{}^{+0.4}_{-0.3}$ & $55.2^{+5.1}_{-5.0}{}^{+0.6}_{-0.6}$ & $60.3^{+4.5}_{-5.5}{}^{+0.7}_{-0.7}$ & $61.9^{+5.0}_{-5.8}{}^{+0.7}_{-0.7}$ \\ \hline
$\gamma W^-$ & nnlo & $16.1^{+1.2}_{-1.3}{}^{+0.3}_{-0.2}$ & $42.1^{+4.3}_{-4.1}{}^{+0.5}_{-0.5}$ & $45.4^{+4.0}_{-4.5}{}^{+0.5}_{-0.6}$ & $46.5^{+4.3}_{-4.8}{}^{+0.5}_{-0.6}$ \\ \hline
$\gamma W^-$ & an$^3$lo & $16.2^{+1.2}_{-1.3}{}^{+0.3}_{-0.2}$ & $42.0^{+4.2}_{-4.1}{}^{+0.4}_{-0.6}$ & $45.3^{+3.9}_{-4.5}{}^{+0.5}_{-0.5}$ & $46.4^{+4.2}_{-4.7}{}^{+0.5}_{-0.6}$ \\ \hline
\end{tabular}
\caption[]{$p p \to \gamma W^+$ and $p p \to \gamma W^-$ cross sections at the LHC with $\sqrt{S}=13$ TeV and MSHT20 pdf. The central results are for $\mu_F=\mu_R=m_W$ and are shown together with scale and pdf uncertainties.}
\label{tableW13teV}
\end{center}
\end{table}

In Table 1, we present the cross sections for $\gamma W^+$ and $\gamma W^-$ production at 13 TeV LHC energy with a cut of 25 GeV on the transverse momentum, $p_T$, of the photon. Results are given at LO, NLO, aNNLO, and aN$^3$LO with scale and pdf uncertainties using MSHT20 nnlo and an$^3$lo pdf. We note that at LO there is no $\mu_R$ uncertainty since that order is $\alpha_s^0$. The magnitude of the scale uncertainty with independent variation of $\mu_F$ and $\mu_R$ is somewhat larger at NLO than at LO; while at aNNLO and aN$^3$LO it is similar to the one at NLO. On the other hand, we note that if we vary the factorization and renormalization scales simultaneously, then the scale uncertainty, at the same perturbative order, is much smaller and is reduced when going to higher orders. For example, at 13 TeV using nnlo pdf for the $\gamma W^+$ cross section with scale uncertainties from simultaneous variation of $\mu_F$ and $\mu_R$, we find values (in pb) of $55.3^{+1.7}_{-1.6}$ (i.e. +3.1\% -2.9\%) at NLO, $60.4^{+1.6}_{-1.5}$ (i.e. +2.6\% -2.5\%) at aNNLO, and $62.0^{+0.8}_{-0.2}$ (i.e. +1.3\% -0.3\% at aN$^3$LO), i.e we see progressively smaller uncertainties at higher orders; the corresponding values (in pb) for the $\gamma W^-$ cross section are $42.1^{+1.5}_{-1.3}$, $45.4^{+1.3}_{-1.1}$, and $46.5^{+0.7}_{-0.0}$ where again the aN$^3$LO uncertainty is smaller than at aNNLO which in turn is smaller than at NLO. Naturally, we expect that the inclusion of hard gluon corrections would affect the numerics of the scale variation. In Table 1, we give the conservative error estimate from independent scale variation but we would expect a reduction in that variation in a complete result that includes hard real radiation at N$^3$LO. 

\begin{table}[htbp]
\begin{center}
\begin{tabular}{|c|c|c|c|c|c|} \hline
\multicolumn{6}{|c|}{$p p \to \gamma W^+$ and $p p \to \gamma W^-$ cross sections at 13.6 TeV \hspace{8mm} $p_{\gamma T}^{\rm cut}=25$ GeV} \\ \hline
process & pdf order &  $\sigma$ LO (pb) & $\sigma$ NLO (pb) & $\sigma$ aNNLO (pb) & $\sigma$ aN$^3$LO (pb)  \\ \hline
$\gamma W^+$ & nnlo & $24.8^{+1.8}_{-2.0}{}^{+0.4}_{-0.3}$ & $58.6^{+5.7}_{-5.5}{}^{+0.5}_{-0.6}$ & $63.9^{+5.1}_{-6.0}{}^{+0.5}_{-0.7}$ & $65.6^{+5.6}_{-6.3}{}^{+0.6}_{-0.7}$ \\ \hline
$\gamma W^+$ & an$^3$lo & $25.1^{+1.7}_{-2.1}{}^{+0.3}_{-0.4}$ & $58.5^{+5.6}_{-5.5}{}^{+0.7}_{-0.7}$ & $63.8^{+5.0}_{-5.9}{}^{+0.7}_{-0.7}$ & $65.5^{+5.6}_{-6.2}{}^{+0.8}_{-0.7}$ \\ \hline
$\gamma W^-$ & nnlo & $17.1^{+1.2}_{-1.5}{}^{+0.2}_{-0.3}$ & $44.9^{+4.6}_{-4.5}{}^{+0.4}_{-0.6}$ & $48.4^{+4.3}_{-4.9}{}^{+0.5}_{-0.6}$ & $49.5^{+4.7}_{-5.2}{}^{+0.5}_{-0.6}$ \\ \hline
$\gamma W^-$ & an$^3$lo & $17.2^{+1.2}_{-1.5}{}^{+0.2}_{-0.3}$ & $44.7^{+4.6}_{-4.4}{}^{+0.5}_{-0.5}$ & $48.2^{+4.3}_{-4.8}{}^{+0.6}_{-0.6}$ & $49.3^{+4.7}_{-5.0}{}^{+0.6}_{-0.6}$ \\ \hline
\end{tabular}
\caption[]{$p p \to \gamma W^+$ and $p p \to \gamma W^-$ cross sections at the LHC with $\sqrt{S}=13.6$ TeV and MSHT20 pdf. The central results are for $\mu_F=\mu_R=m_W$ and are shown together with scale and pdf uncertainties.}
\label{tableW136teV}
\end{center}
\end{table}

In Table 2, we present the cross sections at LO, NLO, aNNLO, and aN$^3$LO (with scale and pdf uncertainties) for $\gamma W^+$ and $\gamma W^-$ production at 13.6 TeV LHC energy, again with a cut of 25 GeV on the $p_T$ of the photon and using MSHT20 nnlo and an$^3$lo pdf. Here the behavior of the scale uncertainties is similar to what was discussed before for the 13 TeV case.

As can be seen from both Tables, the enhancement from the aNNLO and aN$^3$LO contributions is significant. We also note that the NNLO $\delta(s_4)$ terms (which include the virtual corrections) are small compared to the NNLO soft-gluon corrections; their contribution to the total NNLO soft+virtual corrections is less than 20\% at both 13 and 13.6 TeV energies.  

Next, we consider the decay of the $W$ boson into charged leptons and neutrinos. We use MATRIX~\cite{Grazzini:2017mhc} to compute the cross section for $p p \to e^- \bar\nu_e \gamma$ and $p p \to e^+ \nu_e \gamma$ at 13 TeV, including QCD corrections up to NNLO and using MSHT20 nnlo pdf. The cross section is computed in the $G_\mu$-scheme, where the electromagnetic coupling associated with the emission of an on-shell photon is taken to be $\alpha=1/137.036$, while the other electromagnetic couplings are equal to the one defined in terms of the Fermi constant $G_\mu$, namely $\alpha=\sqrt{2}G_\mu m_W^2 (1-m_W^2/m_Z^2)/\pi$. As input parameters we use $m_W=80.3692$ GeV, $m_Z=91.188$ GeV, $G_\mu=1.16638\times 10^{-5}$ GeV$^{-2}$ \cite{RPP24}. Furthermore, the cross section is computed applying some fiducial cuts inspired by a recent experimental analysis \cite{CMS13a}, namely $p_{e T}>25$ GeV, $|\eta_e|<2.5$, $p_{\gamma T}>25$ GeV, $|\eta_\gamma|<2.5$ and $\Delta R(e,\gamma)>0.5$. Here $\Delta R =\sqrt{\Delta\varphi^2+\Delta\eta^2}$, where $\Delta\varphi$ and $\Delta\eta$ are the spatial separations in azimuthal angle $\varphi$ and pseudorapidity $\eta$ between the lepton and photon. The central results are obtained by setting $\mu_F=\mu_R=m_W$, while the scale uncertainties are obtained by varying the factorization and renormalization scales independently. For the process $p p \to e^- \bar\nu_e \gamma$ we find $0.829^{+0.075}_{-0.085}$ pb at LO, $2.39^{+0.26}_{-0.25}$ pb at NLO, and $2.93^{+0.14}_{-0.16}$ pb at NNLO (for simultaneous variation of $\mu_F$ and $\mu_R$, we find $2.39^{+0.08}_{-0.08}$ pb at NLO and $2.93^{+0.12}_{-0.09}$ pb at NNLO). These results show a NLO/LO $K$-factor equal to 2.88 and a NNLO/NLO $K$-factor equal to 1.23. For $p p \to e^+ \nu_e \gamma$ at 13 TeV we find $1.03^{+0.09}_{-0.10}$ pb at LO, $2.75^{+0.28}_{-0.28}$ pb at NLO, and $3.34^{+0.14}_{-0.18}$ pb at NNLO (for simultaneous $\mu_F$ and $\mu_R$ variation, we find $2.75^{+0.09}_{-0.09}$ pb at NLO and $3.34^{+0.12}_{-0.10}$ pb at NNLO). For this process we have a NLO/LO $K$-factor equal to 2.67 and a NNLO/NLO $K$-factor equal to 1.21. We note that the $K$-factors for the processes with decays and the selected cuts are significantly bigger than for those without decays. Assuming that the aN$^3$LO $K$-factors are also bigger for the decay processes, we would infer a minimum value for the aN$^3$LO cross section at 13 TeV of 3.00 pb for $p p \to e^- \bar\nu_e \gamma$ and of 3.43 pb for $p p \to e^+ \nu_e \gamma$. However, since we cannot implement the cuts on the decay particles in our formalism, we cannot make more predictions than this minimum value for the aN$^3$LO cross section. 

\mysection{Photon $p_T$ distributions at 13 and 13.6 TeV}

In this section we present results for the photon transverse-momentum differential distributions in $\gamma W^+$  and $\gamma W^-$ production at the LHC. As for the total cross section, we take $m_W=80.3692$ GeV, $m_Z=91.188$ GeV, and $\alpha^{-1}=127.9$. Here we use an$^3$lo pdf and  present the central results which are obtained by setting $\mu_F=\mu_R=m_W$. Again, the aNNLO results are obtained by adding second-order soft-plus-virtual QCD corrections to the complete NLO result. Third-order soft-gluon corrections are further added in order to derive a result at aN$^3$LO.

\begin{figure}[htbp]
\begin{center}
\includegraphics[width=88mm]{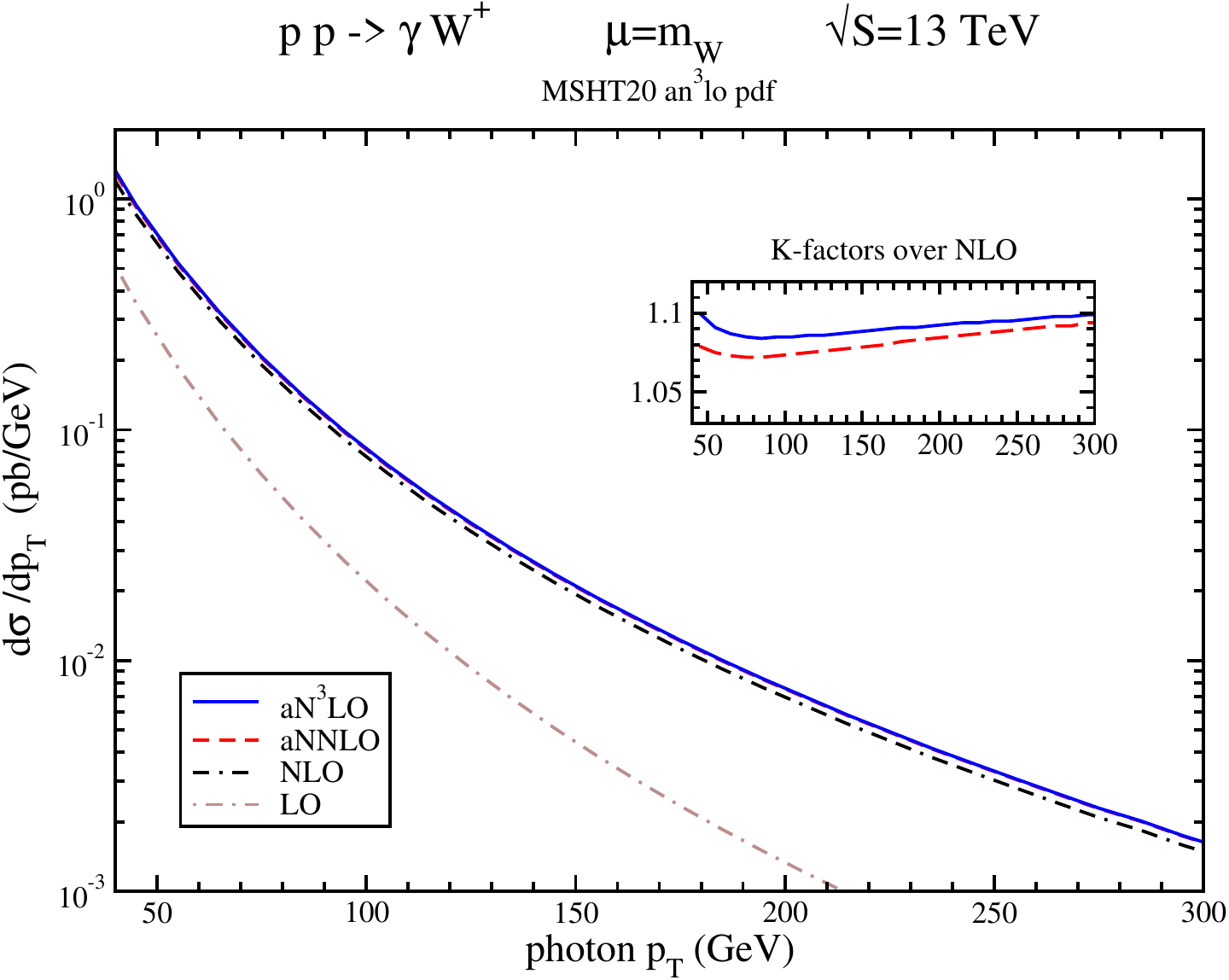}
\includegraphics[width=88mm]{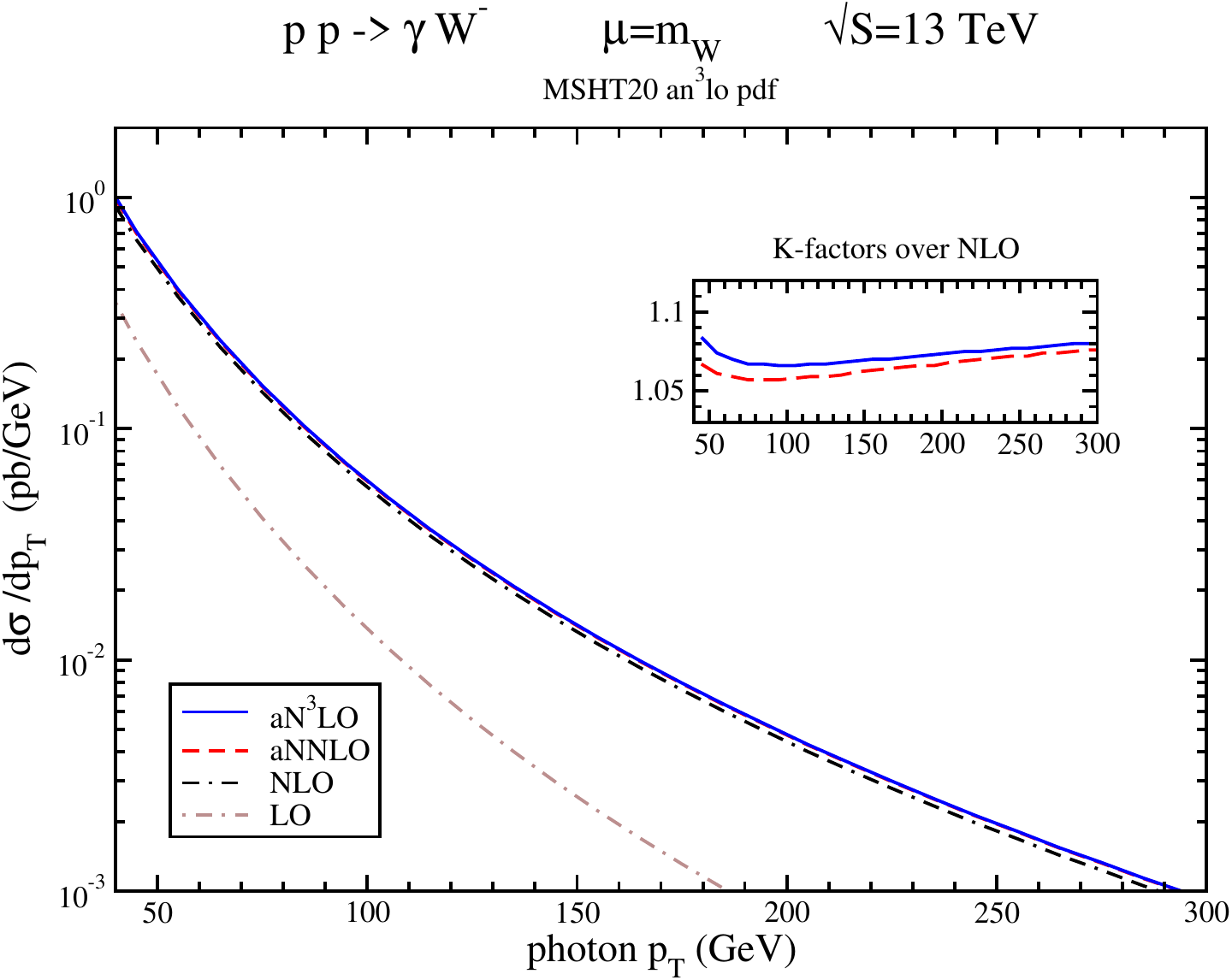}
\caption{The photon $p_T$ distributions through aN$^3$LO in $\gamma W^+$ (left) and $\gamma W^-$ (right) production in $pp$ collisions at an LHC energy of 13 TeV. The inset plots display the $K$-factors relative to NLO.}
\label{ptgammaW13}
\end{center}
\end{figure}

In Fig. \ref{ptgammaW13}, we show the photon $p_T$ distribution in $\gamma W^+$ and $\gamma W^-$ production at 13 TeV LHC energy through aN$^3$LO. The distributions fall quickly by several orders of magnitude as the photon transverse momentum increases. The inset plots show the $K$-factors, i.e. the ratios of the aNNLO and aN$^3$LO results to NLO. The aNNLO and aN$^3$LO distributions are practically on top of each other in the main logarithmic plots, but the inset plots clearly show the additional aN$^3$LO contributions in the $K$-factors. The dependence of the $K$-factors on the photon transverse momentum is mild. The $K$-factors in both plots take values between 1.05 and 1.1 in the $p_T$ range shown, depending on the process and the perturbative order, they have a minimum at a $p_T$ value of around 80 GeV, and they increase slowly at higher $p_T$ values.

\begin{figure}[htbp]
\begin{center}
\includegraphics[width=88mm]{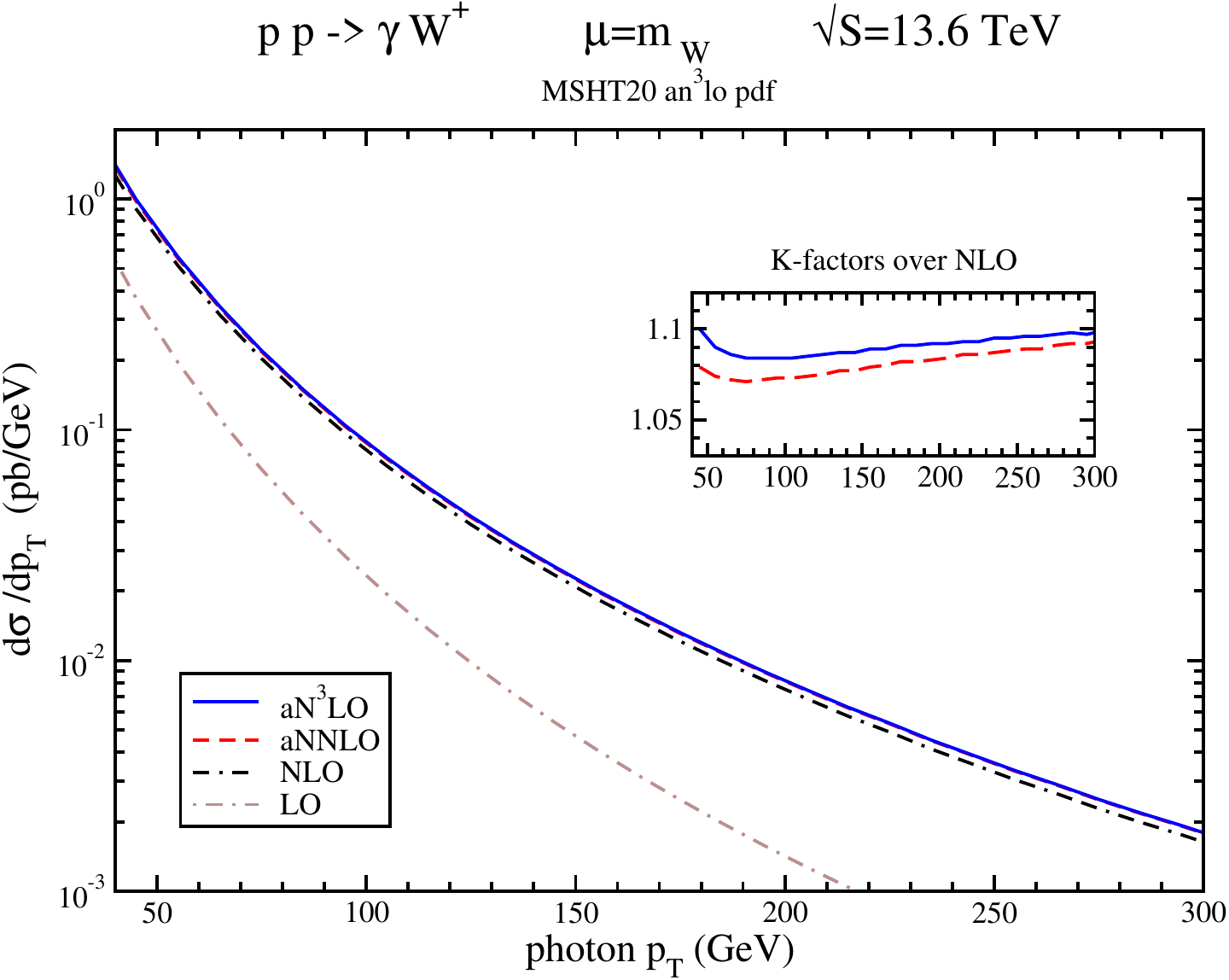}
\includegraphics[width=88mm]{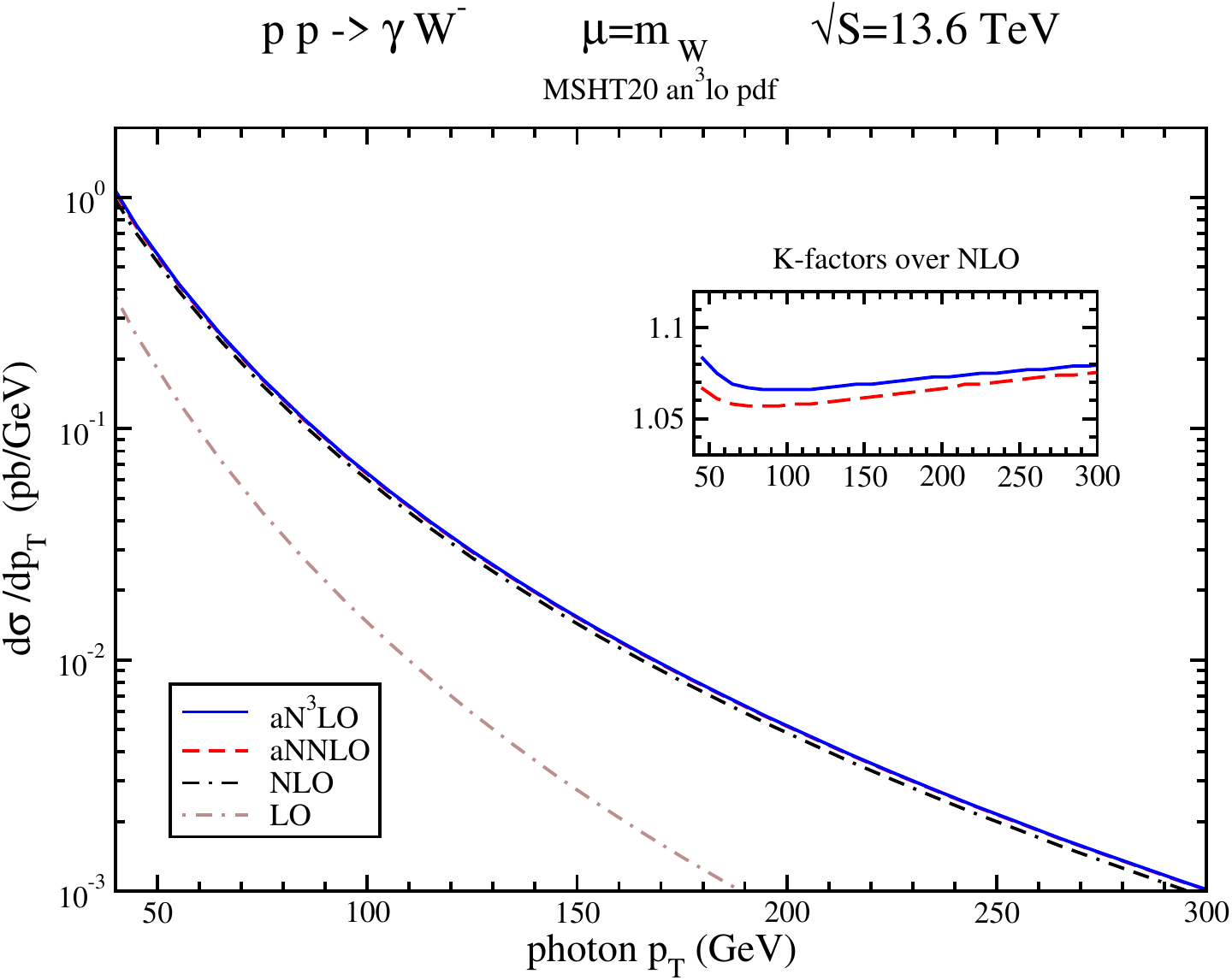}
\caption{The photon $p_T$ distributions through aN$^3$LO in $\gamma W^+$ (left) and $\gamma W^-$ (right) production in $pp$ collisions at an LHC energy of 13.6 TeV. The inset plots display the $K$-factors relative to NLO.}
\label{ptgammaW136}
\end{center}
\end{figure}

In Fig. \ref{ptgammaW136}, we show the photon $p_T$ distribution in $\gamma W^+$ and $\gamma W^-$ production at 13.6 TeV LHC energy through aN$^3$LO. Similarly to the 13 TeV case, we see that the distributions fall very quickly with increasing photon $p_T$. The inset plots show the aNNLO and aN$^3$LO $K$-factors relative to NLO. The $K$-factors at 13.6 TeV are slightly smaller than those at 13 TeV, but the dependence on the transverse momentum is, again, mild. Indeed, they also reach a minimum value at $p_T$ of around 80 GeV, and then they slowly increase at higher $p_T$.

Finally, we note that the shape of the photon $p_T$ distributions as well as the $K$-factors as functions of $p_T$ can be affected by the inclusion of fiducial cuts on the $W$ boson decay products.

\mysection{Conclusions}

In this paper we have provided a study of higher-order QCD corrections to $\gamma W$ production in proton-proton collisions at the LHC, considering a center-of-mass energy of 13 TeV and 13.6 TeV. We have computed aNNLO QCD cross sections for both $\gamma W^+$ and $\gamma W^-$ production by adding the NNLO soft-plus-virtual QCD corrections to the complete NLO result. Our calculation showed that, at this order, soft-gluon corrections numerically dominate the virtual ones. Moreover, results at aN$^3$LO have been derived by including third-order soft-gluon corrections. For the calculation of the total cross section, scale and pdf uncertainties have been also provided. From our results, we obtained that the relative uncertainties from factorization and renormalization scale variation do not change much at higher orders when performing an independent variation of the two scales over the range $m_W/2$ to $2m_W$, while they decrease significantly when considering a simultaneous variation of these scales.  

We have also computed aNNLO and aN$^3$LO differential distributions in the transverse momentum of the photon at 13 and 13.6 TeV LHC energies, and we have presented separate results for the $\gamma W^+$ and $\gamma W^-$ processes. The distributions fall very quickly with increasing $p_T$ of the photon; indeed, they decrease by several orders of magnitude over the $p_T$ range in the plots. The $K$-factors relative to NLO show a mild dependence on the transverse momentum of the photon for both energies and processes, and they slowly increase at higher photon $p_T$ values. Similarly to the total cross section, we find significant enhancements from the higher-order corrections.

This work is the first calculation of complete soft-gluon corrections at N$^3$LO in single-particle-inclusive kinematics for a Standard Model process. Our results show that the enhancements to the total cross sections and the photon $p_T$ distributions in $\gamma W$ production at LHC energies from the inclusion of aNNLO and aN$^3$LO contributions are significant.

\section*{Acknowledgements}
This material is based upon work supported by the National Science Foundation under Grant No. PHY 2412071.

\end{document}